\documentclass[twocolumn,showpacs,preprintnumbers,amsmath,amssymb,nofootinbib,superscriptaddress]{revtex4-1}
\usepackage{amsmath}
\usepackage{amssymb}
\usepackage{subfigure}
\usepackage{times}
\usepackage{graphicx} 

\begin{document}

\title{Modeling and Predicting Popularity Dynamics 
via Reinforced Poisson Processes}
\author{Hua-Wei Shen}
\email{shenhuawei@ict.ac.cn}
\affiliation{Institute of Computing Technology, Chinese Academy of
Sciences, Beijing 100190, China}
\author{Dashun Wang}
\email{dashunwang@gmail.com}
\affiliation{IBM Thomas J. Watson Research Center, Yorktown Heights, New York 10598, USA}
\author{Chaoming Song}
\email{chaoming.song@gmail.com}
\affiliation{Department of Physics, University of Miami, Coral Gables, FL 33146, USA}
\author{Albert L\'{a}szl\'{o} Barab\'{a}si}
\email{barabasi@gmail.com}
\affiliation{Center for Complex Network Research and Departments of Physics, Computer Science and Biology, Northeastern University, Boston, Massachusetts 02115, USA}

\date{\today}

\begin{abstract}
An ability to predict the popularity dynamics of \emph{individual} items within a complex evolving system has important implications in an array of areas. Here we propose a generative probabilistic framework using a reinforced Poisson process to model explicitly the process through which individual items gain their popularity. This model distinguishes itself from existing models via its capability of modeling the arrival process of popularity and its remarkable power at predicting the popularity of individual items. It possesses the flexibility of applying Bayesian treatment to further improve the predictive power using a conjugate prior. Extensive experiments on a longitudinal citation dataset demonstrate that this model consistently outperforms existing popularity prediction methods.
\end{abstract}

\pacs{89.75.Fb, 05.10.-a}

\maketitle

\section{Introduction}
\label{introduction}

The explosive growth of information, from knowledge database to online media, places attention economy in the center of this era. In the heart of attention economy lies a competing process through which a few items become popular while most are forgotten over time~\cite{Wu2007}. For example, videos on YouTube or stories on Digg gain their popularity by striving for views or votes~\cite{Szabo2010}; papers increase their visibility by competing for citations from new papers~\cite{Wang2013}; tweets or Hashtags in Twitter become more popular as being retweeted~\cite{Hong2011} and so do webpages as being attached by incoming hyperlinks~\cite{Ratkiewicz2010}. An ability to predict the popularity of \emph{individual} items within a dynamically evolving system not only probes our understanding of complex systems, but also has important implications in a wide range of domains, from marketing and traffic control to policy making and risk management. Despite recent advances of empirical methods, we lack a general modeling framework to predict the popularity of \emph{individual} items within a complex evolving system.

\begin{figure}[tb]
\vskip 0.2in
\begin{center}
\subfigure[Citations]{
    \label{fig3a}
    \includegraphics[width = 0.45 \textwidth]{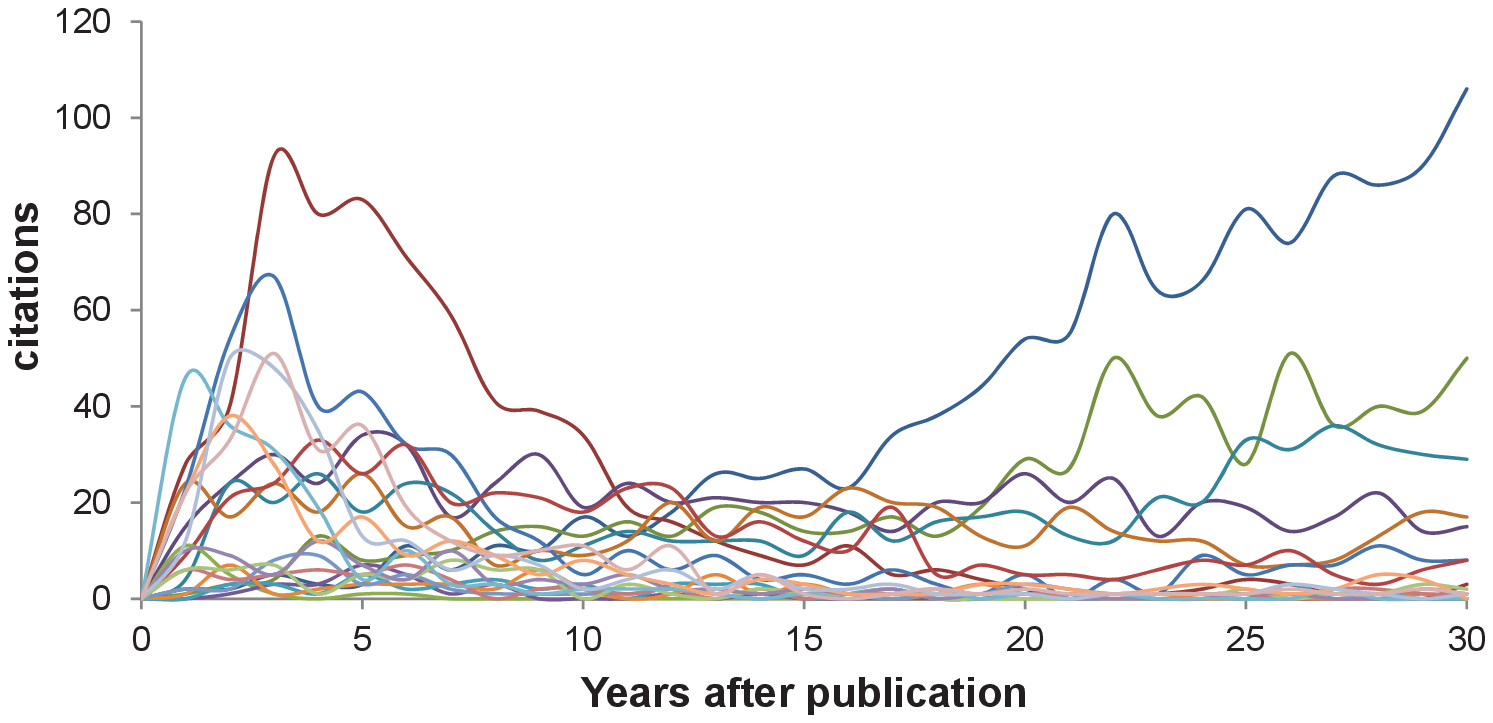}
}
\subfigure[Hashtags]{
    \label{fig3b}
    \includegraphics[width = 0.45 \textwidth]{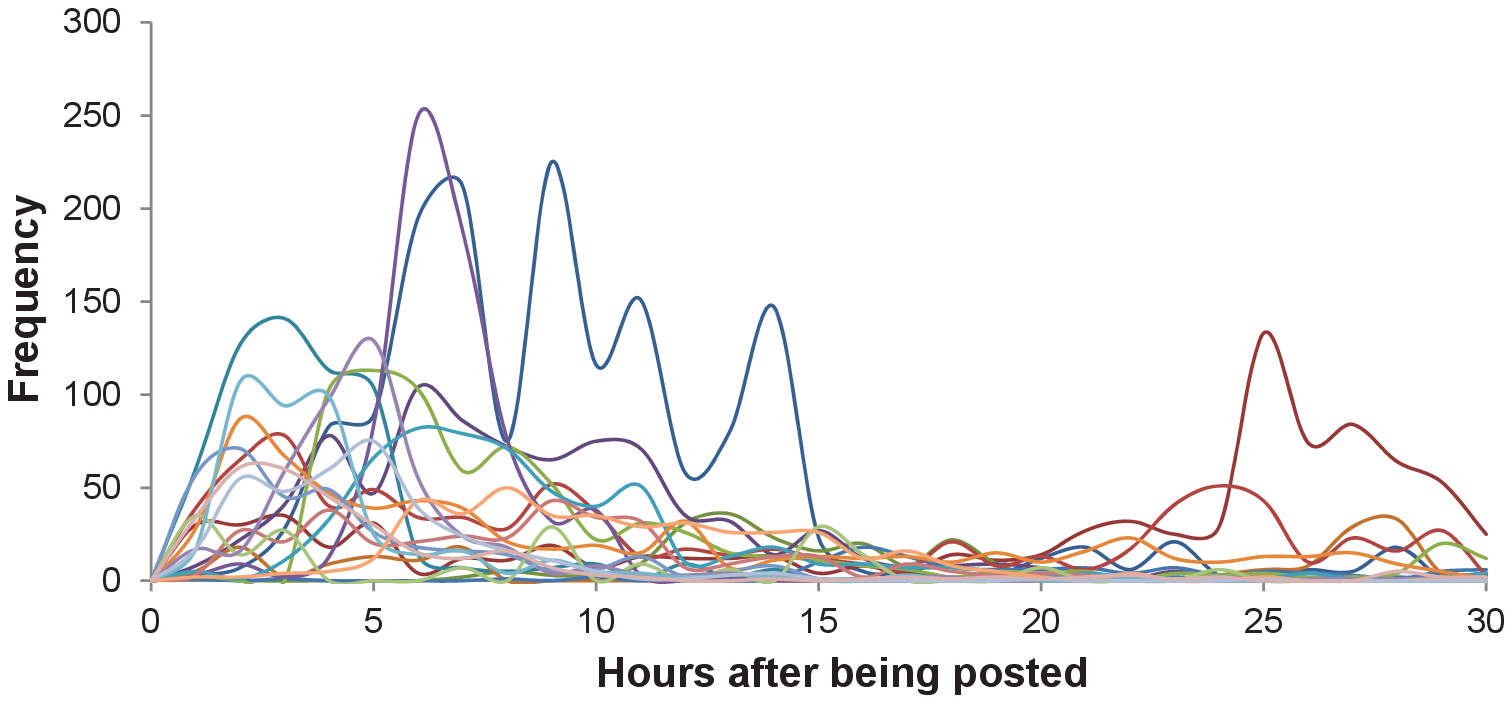}
}
\vskip -0.1in
\caption{\label{fig0} Stochastic Popularity dynamics. (a) Citations of $20$ papers selected randomly from Physical Review during 1960s. (b) Frequency of $20$ Hashtags selected randomly from Twitter in $2012$. }
\end{center}
\vskip -0.2in
\end{figure}

Indeed, current models fall into two main paradigms, each with known strengths and limitations. One focuses on reproducing certain statistical quantities over an aggregation of items~\cite{Barabasi2005,Crane2008,Ratkiewicz2010}. These models have been successful in understanding the underlying mechanisms of popularity dynamics, such as the disparity in popularity. Yet, as they do not provide a way to extract item-specific parameters, these models lack predictive power for the popularity dynamics of individual items. The other line of enquiry, in contrast, treats the popularity dynamics as time series, making predictions by either exploiting temporal correlations~\cite{Szabo2010,Yang2010,Lerman2010,Bao2013b} or fitting to these time series certain classes of functions~\cite{Bass1969,Mahajan1990,Matsubara2012,Gomez-Rodriguez2013}.
Despite their initial success in certain domains, these models are deterministic, modeling the popularity dynamics in a mean-field, if heuristic, fashion by focusing on the \emph{average} amount of attention received within a fixed time window, ignoring the underlying arrival process of attentions. Indeed, to best of our knowledge, we lack a probabilistic framework currently to model and predict the popularity dynamics of individual items. The reason behind this is partly illustrated in Figure~\ref{fig0}, suggesting that the dynamical processes governing individual items appear too noisy to be amendable to quantification.

In this paper, we model the stochastic popularity dynamics using reinforced Poisson processes, capturing simultaneously three key ingredients: Fitness of an item, characterizing its inherent competitiveness against other items; a general temporal relaxation function corresponding to the aging in the ability to attract new attention, and a reinforcement mechanism documenting the well-known ``rich-get-richer'' phenomenon. The benefit of the proposed model is three-fold: (1) It models the arrival process of individual attentions directly in contrast to relying on aggregated popularity time series; (2) As a generative probabilistic model, it can be easily incorporated into the Bayesian framework to account for external factors, hence leading to improved predictive power; (3) The flexibility in its choice of specific relaxation functions makes it a general framework that can be adapted to model the popularity dynamics in different domains.

Taking citation system as an exemplary case, we demonstrate the effectiveness of the proposed framework using a dataset peculiar in its longitudinality, spanning over 100 years and containing all the papers ever published by American Physical Society. We find the proposed model consistently outperforms competing methods. Moreover, the proposed model is general. Hence it is not limited to predicting citations, but with appropriate adjustments will likely apply to other domains driven by competing processes.

\section{Reinforced Poisson Process}
\label{model}

\subsection{Model Formulation}

The popularity dynamics of individual item $d$ during time period $[0,T]$ is characterized by a set of time moments $\{t_i^d\} (1\leq i \leq n_d)$ when each attention is received, where $n_d$ represents the total number of attentions. Without loss of generality, we have $0=t_0^d \leq t_1^d  \leq \cdots \leq t_i^d \leq \cdots \leq t_{n_d}^d \leq T$. To model the arrival process of $\{t_i^d\}$, we consider two major phenomena confirmed independently in previous studies of population dynamics: (1) the reinforcement capturing the ``rich-get-richer'' mechanism, i.e., previous attention triggers more subsequent attentions~\cite{Crane2008}; (2) the aging effect characterizing time-dependent attractiveness of individual items. Taken these two factors together, for an individual item $d$, we model its popularity dynamics as a reinforced Poisson process (RPP)~\cite{Pemantle2007} characterized by the rate function $x_d(t)$ as  \begin{equation}
x_d(t)=\lambda_d f_d(t;\theta_d) i_d(t), \label{eq1}
\end{equation}
where $\lambda_d$ is the intrinsic attractiveness, $f_d(t;\theta_d)$
is the relaxation function that characterizes the temporal inhomogeneity due to the aging effect modulated by parameters $\theta_d$, and $i_d(t)$ is the total number of attentions received up to time $t$. From Bayesian viewpoint, the total number of attentions $i_d(t)$ is the sum of the number of real attentions and the \textit{effective} number of attentions which plays the role of prior belief. Here, we assume that all items are created equal and hence the effective number of attentions for all items has the same value, denoted by $m$. Therefore during the time interval between the $(i-1)$th and $i$th attentions, we have
\begin{equation}
i_d(t) = m+i-1, \label{eq2}
\end{equation}
where $1 \leq i \leq n_d$. Accordingly, during the time interval between the $n_d$th attention and $T$, the total number of attention is $m+n_d$.

\begin{figure}[t]
\vskip 0.1in
\begin{center}
\centerline{\includegraphics[width=0.4\textwidth]{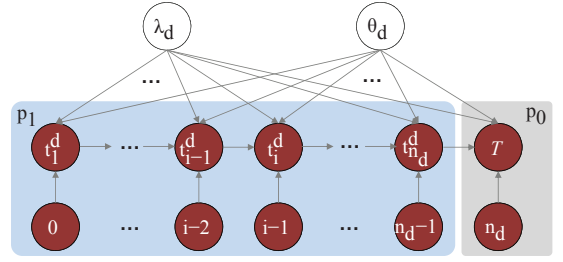}}
\caption{Graphical representation of the generative model for popularity dynamics via reinforced Poisson process.}
\label{fig1}
\end{center}
\vskip -0.2in
\end{figure}

The length of time interval between two consecutive attentions follows an inhomogeneous Poisson process. Therefore, given that the $(i-1)$th attention arrives at $t_{i-1}^d$, the probability that the $i$th attention arrives at $t_i^d$ follows
\begin{eqnarray}
p_1(t_i^d| t_{i-1}^d) &=& \lambda_d f_d(t_i^d;\theta_d) (m+i-1) \nonumber \\
&& \times e^{-\int_{t_{i-1}^d}^{t_i^d} \lambda_d f_d(t;\theta_d) (m+i-1) \mbox{d}t}, \label{eq3}
\end{eqnarray}
and the probability that no attention arrives between $t_{n_d}^d$ and $T$ is
\begin{eqnarray}
p_0(T|t_{n_d}^d) &=& e^{-\int_{t_{n_d}^d}^{T} \lambda_d f_d(t;\theta_d) (m+n_d) \mbox{d}t}. \label{eq4}
\end{eqnarray}

Incorporating Eqs.~(\ref{eq3}) and (\ref{eq4}) with the fact that attentions during different time intervals are statistically independent, the likelihood of observing the popularity dynamics $\{t_i^d\}$ during time interval $[0,T]$ follows
\begin{eqnarray}
\mathcal{L}(\lambda_d,\theta_d) &=& p_0(T|t_{n_d}^d)\prod_{i=1}^{n_d}{p_1(t_i^d| t_{i-1}^d)} \nonumber \\
&=& \lambda_d^{n_d}\prod_{i=1}^{n_d}{(m+i-1)f_d(t_i^d;\theta_d)} \times \nonumber \\
&&e^{-\lambda_d\left((m+n_d)F_d(T;\theta_d)-\sum_{i=1}^{n_d}{F_d(t_i^d;\theta_d)}\right)},\nonumber \\\label{eq5}
\end{eqnarray}
where $F_d(t;\theta_d)\equiv\int_{0}^{t} f_d(t;\theta_d)  \mbox{d}t$ and we have reorganized the terms on the exponent for simplicity. For clarity, we illustrate the proposed RPP model in the graphical representation (Figure~\ref{fig1}).

\subsection{Parameter Estimation and Prediction}
By maximizing the likelihood function in Eq.~(\ref{eq5}), we obtain the most likely fitness parameter $\lambda_d^*$ for item $d$ in closed form:
\begin{equation}
\lambda_d^* = \frac{n_d}{(m+n_d)F_d(T;\theta_d^*)-\sum_{i=1}^{n_d}{F_d(t_i^d;\theta_d^*)} }. \label{eq6}
\end{equation}
The solution for $\theta_d^*$ depends on the specific form of relaxation function $f_d(t;\theta_d)$. We save the discussions about the estimation of $\theta_d^*$ for later.

Next we show that, with the obtained $\lambda_d^*$ and $\theta_d^*$, the model can be used to predict the expected number $c^d(t)$ of attention gathered by item $d$ up to any given time $t$. Indeed, according to Eq.~(\ref{eq1}), for $t\geq T$, this prediction task is equivalent to the following differential equation
\begin{equation}
\frac{\mbox{d}c^d(t)}{\mbox{d}t} = \lambda_d f_d(t;\theta_d)(m+c^d(t)) \label{eq7}
\end{equation}
with the boundary condition $c^d(T)=n_d$. Solving this differential equation, we get the prediction function
\begin{equation}
c^d(t) = (m+n_d)e^{\lambda_d^* \big(F_d(t;\theta_d^*)-F_d(T;\theta_d^*)\big)} - m. \label{eq8}
\end{equation}

\section{Reinforced Poisson Process with prior}

Maximum likelihood parameter estimation suffers from overfitting problem for small sample size. For example, Eq.~(\ref{eq6}) gives $\lambda_d^*=0$ when $n_d=0$, and results in a null forecasting of future popularity, i.e., the expected number $c^d(t)$ of attention is $0$ at any future time $t$. Moreover, the exponential dependency of $c^d(t)$ on $\lambda_d^*$ in Eq.~(\ref{eq8}) leads to a large uncertainty in the prediction of $c^d(t)$. In this section, to overcome the drawback of the parameter estimation in Eq.~(\ref{eq6}), we adopt the Bayesian treatment for popularity prediction by introducing a conjugate prior for the fitness parameter $\lambda_d$, leading to a further improvement of the prediction accuracy of the proposed RPP model.

\subsection{Conjugate Prior}

\begin{figure}[t]
\begin{center}
\centerline{\includegraphics[width=0.5\columnwidth]{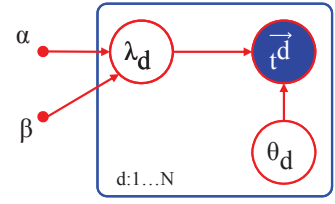}}
\caption{Probabilistic graphical model for popularity dynamics with conjugate prior.}
\label{fig2}
\end{center}
\vskip -0.2in
\end{figure}

The likelihood function, as shown in Eq.~(\ref{eq5}), is a product of a power function and an exponential function of $\lambda_d$. Therefore, the conjugate prior for $\lambda_d$ follows the gamma distribution
\begin{equation}
p(\lambda_d|\alpha,\beta) = \frac{\beta^\alpha}{\Gamma(\alpha)}\lambda_d^{\alpha-1}e^{-\beta\lambda_d}. \label{eq9}
\end{equation}
Note that this conjugate prior is the prior distribution of fitness parameters for all $N$ items rather than for certain individual item. Hereafter, for convenience, we use $\vec{t^d}\equiv \{t_i^d\}$ to denote all the arrival time of attention gathered by item $d$. After introducing the conjugate prior, the graphical representation of model is depicted in Figure~\ref{fig2}.

Using Bayes' theorem and combining Eqs.~(\ref{eq5}) and (\ref{eq9}), we obtain the posterior distribution of $\lambda_d$
\begin{eqnarray}
p(\lambda_d | \vec{t^d}, \theta_d, \alpha, \beta) &=& \frac{p(\vec{t^d}|\lambda_d, \theta_d)p(\lambda_d | \alpha, \beta)}{\int{p(\vec{t^d}|\lambda_d, \theta_d)p(\lambda_d | \alpha, \beta)}\mbox{d}\lambda_d} \nonumber
\end{eqnarray}
\begin{equation}
=\frac{(\beta+X)^{\alpha+n_d}}{\Gamma(\alpha+n_d)}\lambda_d^{\alpha+n_d-1}e^{-(\beta+X)\lambda_d}, \label{eq10}
\end{equation}
where $X\equiv (m+n_d)F_d(T;\theta_d)-\sum_{i=1}^{n_d}{F_d(t_i^d;\theta_d)}$. Comparing Eq.~(\ref{eq9}) and Eq.~(\ref{eq10}), we can see that the form of posterior distribution is the same to the form of prior distribution, reflecting the benefit of conjugate prior. The main difference between posterior distribution and prior distribution lies in the change of parameter values, mediated by the likelihood function.

\subsection{Bayesian Prediction}

With the obtained posterior distribution of $\lambda_d$, the expected number of attention $c^d(t)$, as shown in Eq.~(\ref{eq8}), can be predicted using its mean over the posterior distribution as
\begin{eqnarray}
\langle c^d(t)\rangle &=& \int{c^d(t) p(\lambda_d | \vec{t^d}, \theta_d, \alpha, \beta)}\mbox{d}\lambda_d \nonumber \\
&=& (m+n_d)\left(\frac{\beta+X}{\beta+X-Y}\right)^{\alpha+n_d}-m,\label{eq11}
\end{eqnarray}
where $Y\equiv F_d(t;\theta_d)-F_d(T;\theta_d)$. Eq.~(\ref{eq11}) is the Bayesian prediction function, predicting $c^d(t)$ using the posterior distribution of $\lambda_d$ instead of using a single value of $\lambda_d^*$ obtained by maximum likelihood estimation. In addition, neither $X$, corresponding to empirical observations, nor $Y$, reflecting the rate difference in reinforced Poisson process, is in the exponent, indicating the robustness of this prediction function. Lastly, using the posterior distribution of $\lambda_d$, we calculate the variance of $c^d(t)$ as the confidence of prediction, obtaining
\begin{eqnarray}
\mbox{var}(c^d(t)) &=& \int{(c^d(t))^2 p(\lambda_d | \vec{t^d}, \theta_d, \alpha, \beta)}\mbox{d}\lambda_d - \langle c^d(t)\rangle^2\nonumber \\
&=& (m+n_d)^2\left[\left(\frac{\beta+X}{\beta+X-2Y}\right)^{\alpha+n_d}-\right.\nonumber \\
&&\left.\left(\frac{\beta+X}{\beta+X-Y}\right)^{2(\alpha+n_d)}\right]. \label{eq12}
\end{eqnarray}
We will, in Section~\ref{experiments}, compare the Bayesian prediction in Eq.~(\ref{eq11}) to the one without prior in Eq.~(\ref{eq8}) through extensive experiments on real dataset.

\subsection{Parameter estimation}

We now discuss how to determine the parameters $\alpha$ and $\beta$ of prior distribution. Basically, the values of prior parameters could be tuned by checking the accuracy of prediction function with respect to prior parameters on so-called validation set. This means that we need to know the future popularity of some items to determine prior parameters. It is impractical in many scenarios where it is unrealistic to leverage future information for prediction. In addition, this method is usually time-consuming since the model has to be trained many times during the process of tuning prior parameters.

One alternative solution is the fully Bayesian approach which introduces hyperprior for prior parameters. Although fully Bayesian approach is theoretically elegant, the inference of prior parameters is intractable in most cases. Approximation methods or Monte Carlo methods have to be adopted.
As a result, the benefit of fully Bayesian approach is discounted by approximation gap in approximation methods or high computational cost of Monte Carlo methods.

In this paper, we determine the value of prior parameters by adopting maximum likelihood estimation with latent variable. Specifically, we choose the $\alpha$ and $\beta$ values that maximize the following logarithmic likelihood function
\begin{eqnarray}
\mathcal{L}(\alpha,\beta) &=& \sum_{d=1}^{N}{ \ln \int { p(\vec{t^d}|\lambda_d)p(\lambda_d | \alpha, \beta)}\mbox{d}\lambda_d}. \label{eq13}
\end{eqnarray}
Here, $\theta_d$ is not explicitly written to keep the notation uncluttered. In sum, $\alpha$ and $\beta$ are obtained according to
\begin{eqnarray}
\frac{\partial \mathcal{L}(\alpha,\beta)}{\partial \beta} &=& \frac{N\alpha}{\beta}-\sum_{i=1}^{N}\lambda_d, \label{eq14} \\
\frac{\partial \mathcal{L}(\alpha,\beta)}{\partial \alpha} &=& N(\ln\beta -\phi_0(\alpha)) + \sum_{d=1}^{N}{ \ln\frac{\lambda_d}{\alpha+n_d} } \nonumber \\
&& + \sum_{d=1}^{N}{ \phi_0(\alpha+n_d)}, \label{eq15}
\end{eqnarray}
where $\phi_0$ is the digamma function and the latent variable is
\begin{eqnarray}
\lambda_d = \frac{\alpha+n_d}{\beta+(m+n_d)F_d(T;\theta_d)-\sum_{i=1}^{n_d}{F_d(t_i^d;\theta_d)} }. \label{eq16}
\end{eqnarray}
Comparing Eq.~(\ref{eq16}) and Eq.~(\ref{eq6}), we can see that the fitness parameter $\lambda_d$ is adjusted by prior parameters $\alpha$ and $\beta$.

Note that the parameters $\theta_d$ for all items are also determined by maximizing the likelihood function in Eq.~(\ref{eq13}). The calculation depends on the specific form of relaxation function $f_d(t;\theta_d)$, which is given in experiments on real dataset.

\begin{table}[t]
\caption{Basic statistics of dataset.}
\label{tab1}
\vskip 0.1in
\begin{center}
\begin{tabular}{lrrcr}
\hline
Journal & \#Papers & \#Citations & Period & \\
\hline
PRSI    &   1, 469 &         668 & 1893-1912 \\
PR      &  47, 941 &    590, 665 & 1913-1969 \\
PRA     &  53, 655 &    418, 196 & 1970-2009 \\
PRB     & 137, 999 & 1, 191, 515 & 1970-2009 \\
PRC     &  29, 935 &    202, 312 & 1970-2009 \\
PRD     &  56, 616 &    526, 930 & 1970-2009 \\
PRE     &  35, 944 &    154, 133 & 1993-2009 \\
PRL     &  95, 516 & 1, 507, 974 & 1958-2009 \\
RMP     &   2, 926 &    115, 697 & 1929-2009 \\
PRSTAB  &   1, 257 &      2, 457 & 1998-2009 \\
PRSTPER &       90 &           0 & 2005-2009 \\
\hline
Total   & 463, 348 & 4, 710, 547 & 1893-2009 \\
\hline
\end{tabular}
\end{center}
\vskip -0.15in
\end{table}

\section{Experiments}
\label{experiments}

In this section, we demonstrate the effectiveness of the proposed RPP model, with and without prior.

\subsection{Experiment setup}

\noindent \textbf{Dataset.} We conduct experiments on an excellent longitudinal dataset, containing all papers and citations from 11 journals of American Physical Society between 1893 and 2009. We choose this dataset for three main reasons: (1) It covers an extended period of time, spanning $117$ years, ideal for modeling and predicting temporal dynamics; (2) treating papers as items, their popularity is relatively well-defined, characterized by citations; (3) all citations are specific to the physics community and thus they reflect the collective behavior of a relatively homogeneous population. Statistics about this dataset is shown in Table~\ref{tab1}.

\noindent \textbf{Relaxation function.} When formalizing the model for popularity dynamics, we introduced a general relaxation function $f_d(t; \theta_d)$ and skipped the discussion of parameter $\theta_d$. Here, when applying this model to a specific case, i.e., to citation system, we need to determine the specific form of the relaxation function as well as $\theta_d$. Previous studies~\cite{Radicchi2008,Wang2013} on citation dynamics suggest that the aging of papers is captured by a log-normal relaxation function
\begin{equation}
f_d(t;\mu_d, \sigma_d) = \frac{1}{\sqrt{2\pi}\sigma_d t}\exp\left(-\frac{(\ln t-\mu_d)^2}{2\sigma_d^2}\right), \label{eq17}
\end{equation}
a common relaxation function, which is also observed in other domains such as messages in microblogging network~\cite{Bao2013a}.

For item $d$ with log-normal relaxation function, $\theta_d$ is replaced by parameters $\mu_d$ and $\sigma_d$, which can be calculated by maximizing the logarithmic likelihood $\mathcal{L}$ in Eq.~(\ref{eq13}) and Eq.~(\ref{eq5}) for the proposed RPP model with and without prior, respectively. In this paper, we maximize logarithmic likelihood using optimization methods which leverage gradients
\begin{eqnarray}
\frac{\partial \mathcal{L}}{\partial \mu_d } &=& \frac{1}{\sigma_d}\left\{ \sum_{i=1}^{n_d}{\Big[\tau_i^d - \lambda_d \phi(\tau_i^d)\Big]} \right. \nonumber \\
&& + \lambda_d(n_d+m)\phi(\tau^d)  \Bigg\}, \label{eq18} \\
\frac{\partial \mathcal{L}}{\partial \sigma_d } &=& \frac{1}{\sigma_d}\left\{ \sum_{i=1}^{n_d}{\Big[\tau_i^d * \tau_i^d - \lambda_d\tau_i^d \phi(\tau_i^d)\Big]} \right. \nonumber \\
&& + \lambda_d(n_d+m)\tau^d \phi(\tau^d) - n_d\Bigg\}, \label{eq19}
\end{eqnarray}
where $\phi$ is the probability density function of standard normal distribution, $\tau_i^d\equiv(\ln t_i^d-\mu_d)/\sigma_d$ and $\tau^d\equiv(\ln T -\mu_d)/\sigma_d$. Therefore, we can use Eqs.~(\ref{eq18}) and (\ref{eq19}) together with Eqs.~(\ref{eq14}) and (\ref{eq15}) to maximize the logarithmic likelihood in Eq.~(\ref{eq13}) for the RPP model with prior, together with Eq.~(\ref{eq6}) to maximize the likelihood in Eq.~(\ref{eq5}). for the RPP model without prior.

\noindent \textbf{Baseline models and evaluation metrics.} To compare the predictive power of the RPP model against other models, we identify three models that have been used or can be used to model and predict popularity dynamics: the classic autoregression (AR) method~\cite{Box2008}, the linear regression method of logarithmic popularity (SH)~\cite{Szabo2010}, and the WSB model~\cite{Wang2013}, which is equivalent to the proposed RPP model without prior when the log-normal relaxation function is adopted. We adopt two standard measurements as evaluation metrics:
\begin{itemize}
\item Mean Absolute Percentage Error (\textit{MAPE}) measures the average deviation between predicted and empirical popularity over an aggregation of items. Denoting with $c^d(t)$ the predicted number of citations for a paper $d$ up to time $t$ and with $r^d(t)$ its real number of citations, we obtain the MAPE over $N$ papers
    \begin{eqnarray}
    MAPE = \frac{1}{N}\sum_{d=1}^{N}\left|\frac{c^d(t)-r^d(t)}{r^d(t)}\right|.\nonumber
    \end{eqnarray}

\item \textit{Accuracy} measures the fraction of papers correctly predicted for a given error tolerance $\epsilon$. Hence the accuracy of popularity prediction on $N$ papers is
    \begin{eqnarray}
    \frac{1}{N}\sum_{d=1}^{N}{|\{d:  \left|\frac{c^d(t)-r^d(t)}{r^d(t)}\right|\leq \epsilon\}|}.\nonumber
    \end{eqnarray}
    We set the threshold $\epsilon=0.1$ in this paper.
\end{itemize}

\begin{figure*}[!htb]
\vskip 0.1in
\begin{center}
\subfigure[Physical Review (1960s)]{
    \label{fig3a}
    \includegraphics[width = 0.30 \textwidth]{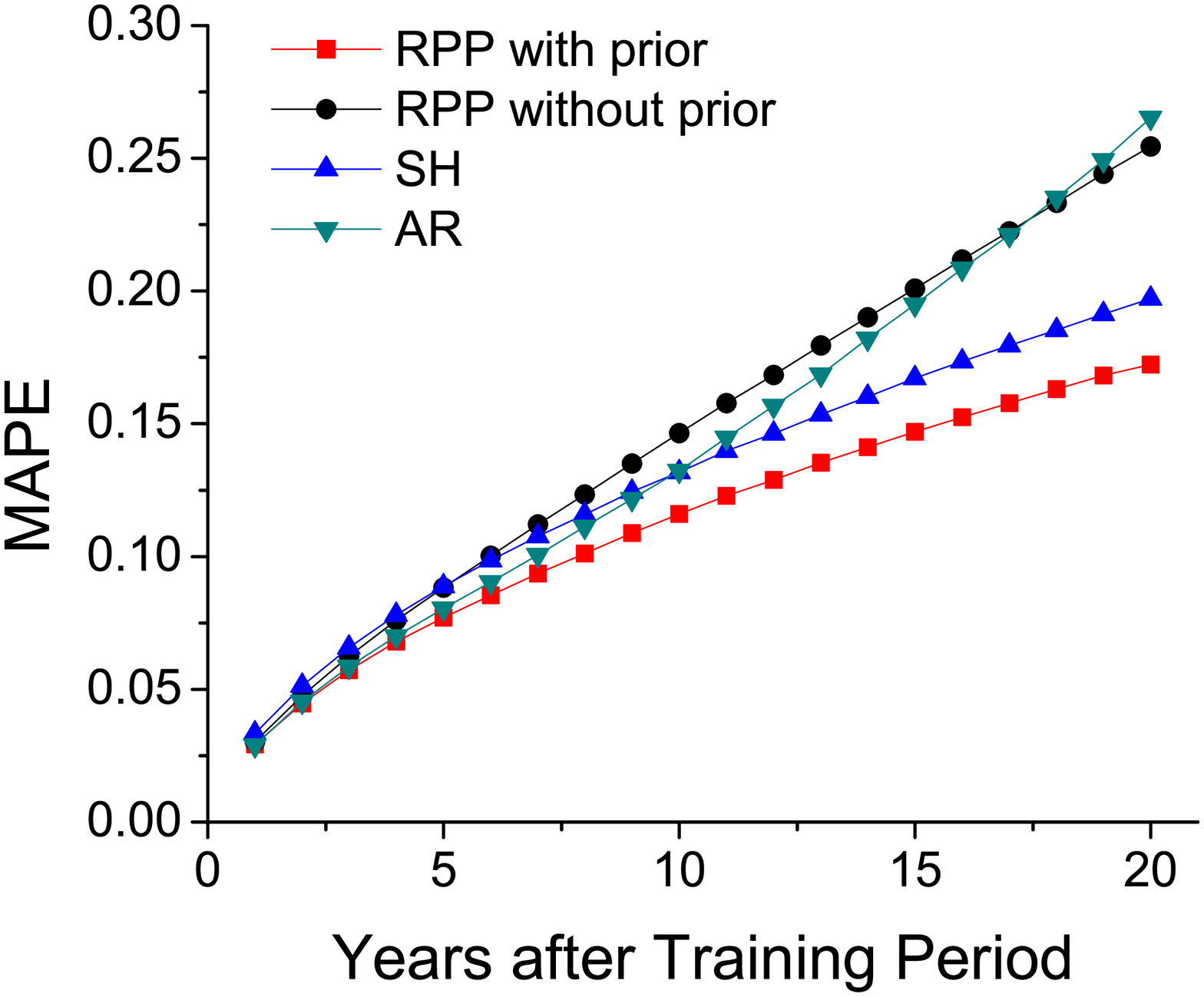}
}
\subfigure[Physical Review Letters (1970s)]{
    \label{fig3b}
    \includegraphics[width = 0.30 \textwidth]{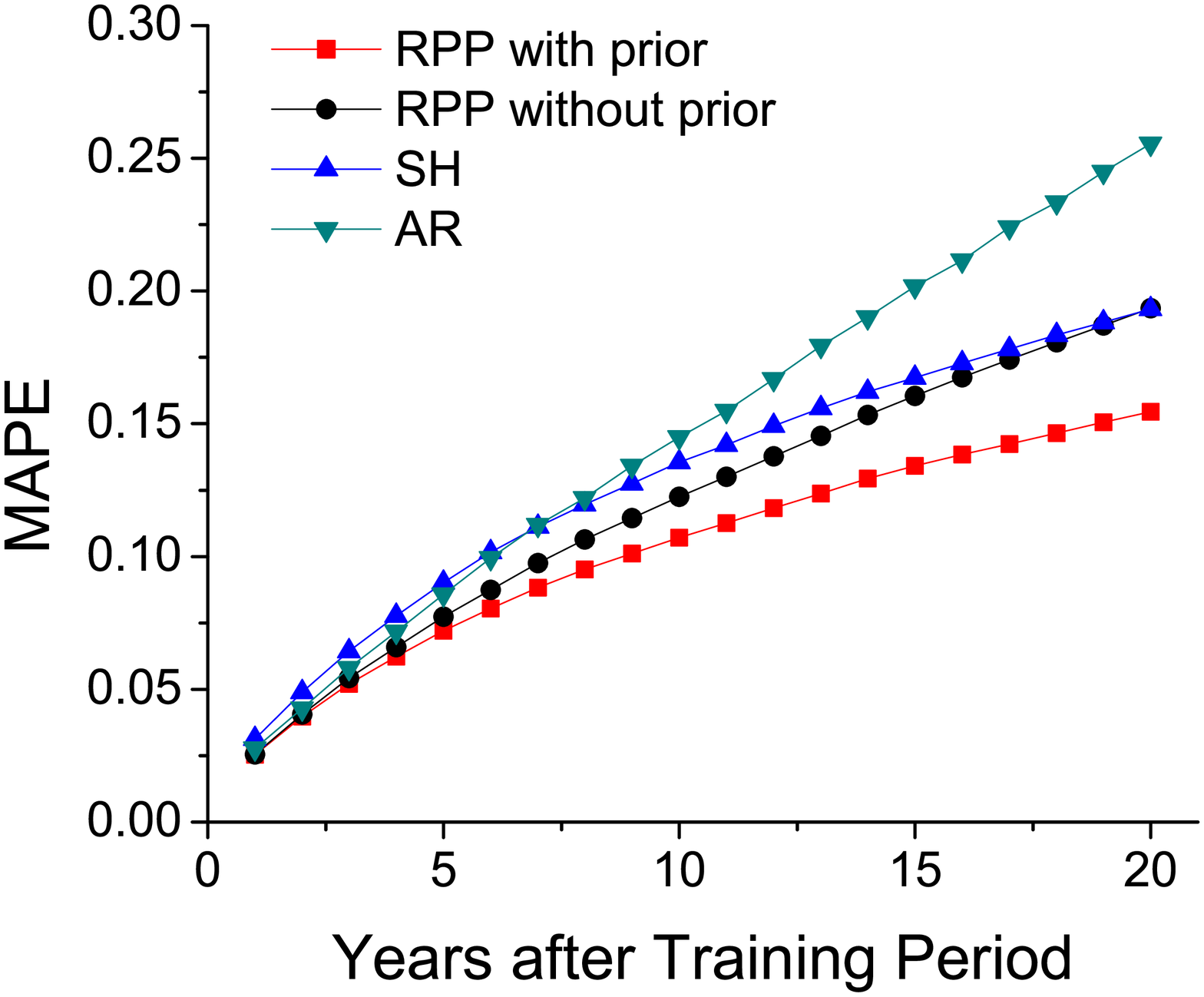}
}
\subfigure[Physical Review B (1980s)]{
    \label{fig3c}
    \includegraphics[width = 0.30 \textwidth]{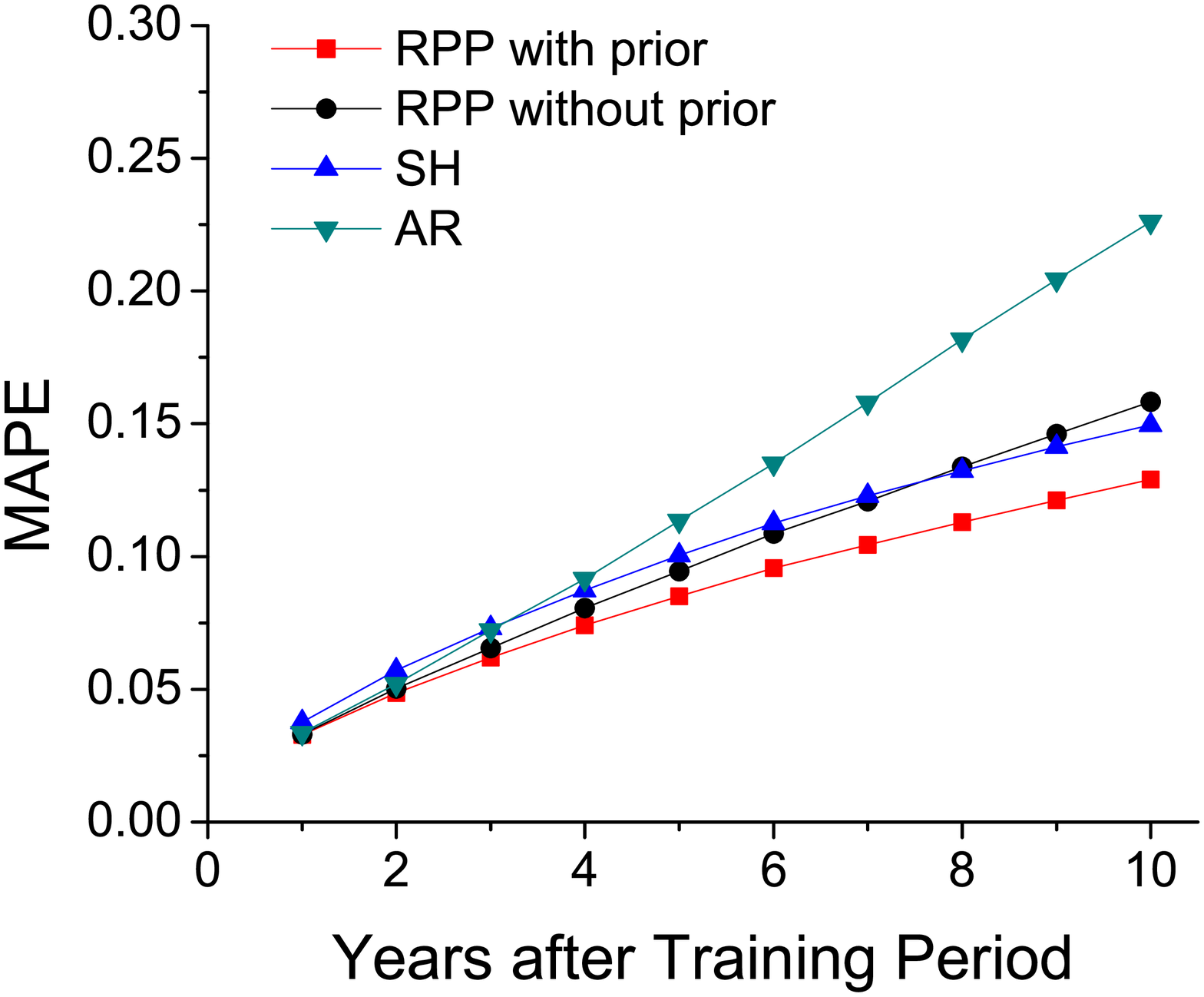}
}
\vskip 0.1in
\subfigure[Physical Review (1960s)]{
    \label{fig3d}
    \includegraphics[width = 0.30 \textwidth]{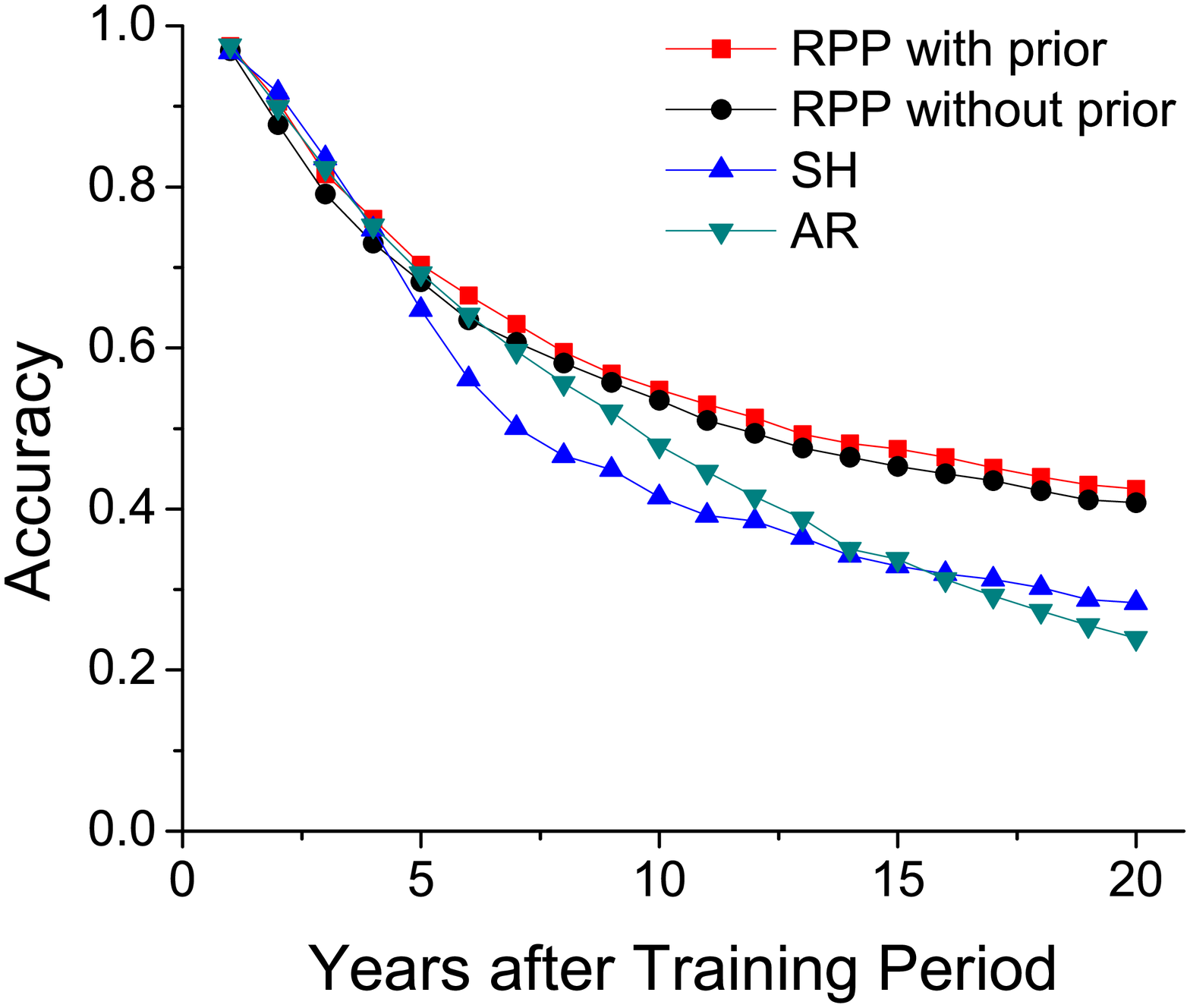}
}
\subfigure[Physical Review Letters (1970s)]{
    \label{fig3e}
    \includegraphics[width = 0.30 \textwidth]{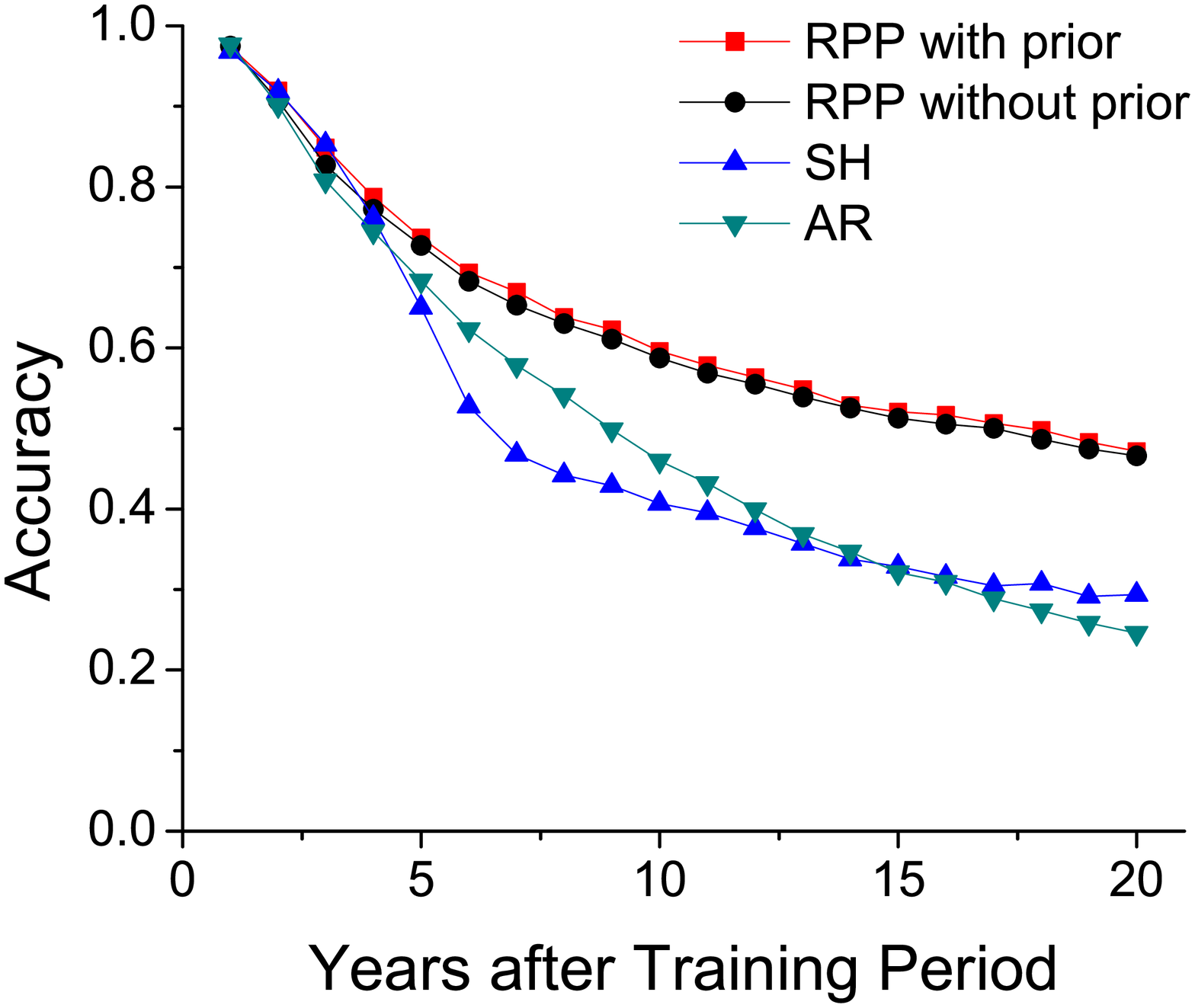}
}
\subfigure[Physical Review B (1980s)]{
    \label{fig3f}
    \includegraphics[width = 0.30 \textwidth]{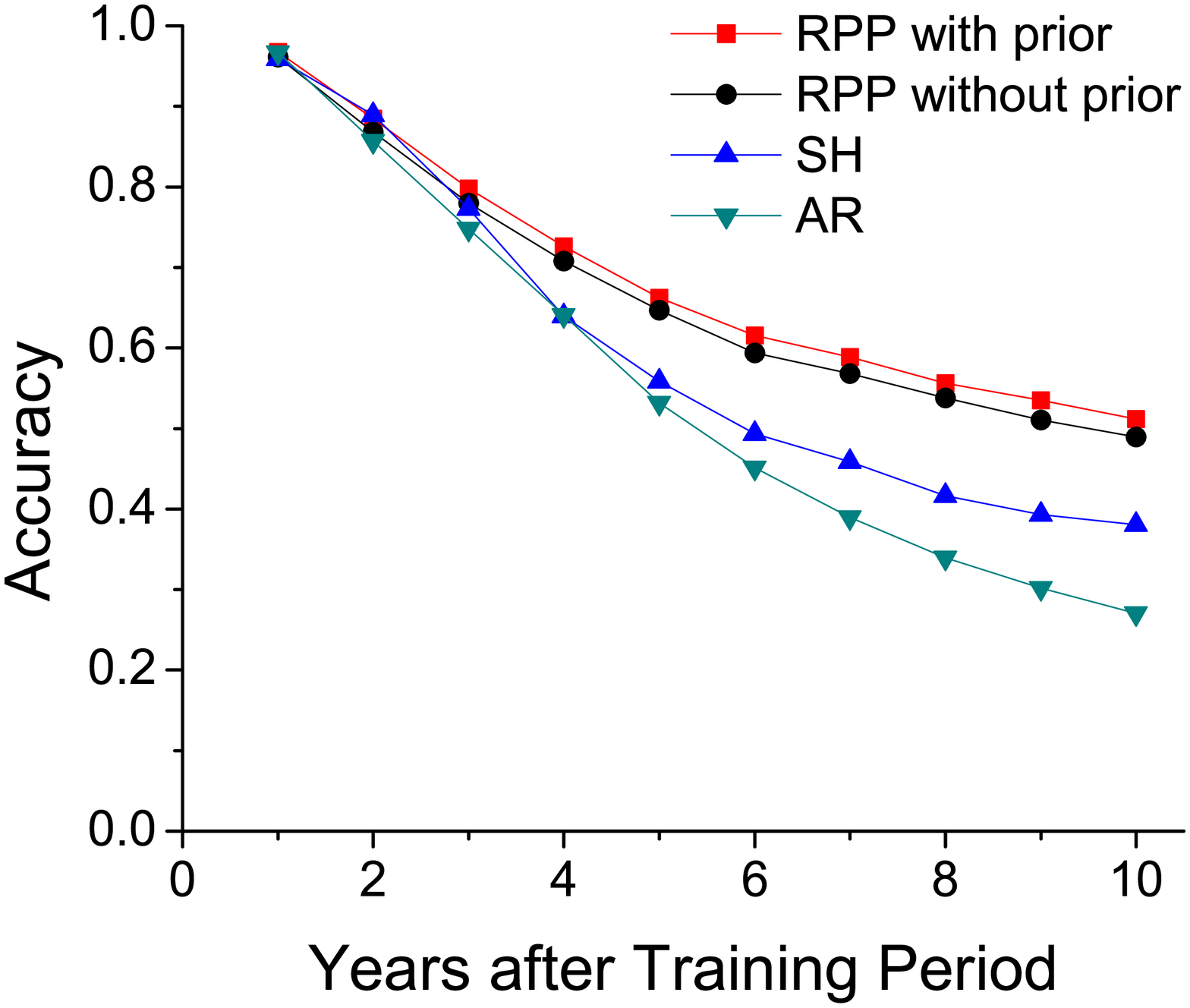}
}
\vskip -0.02in
\subfigure[Physical Review (1960s)]{
    \label{fig3g}
    \includegraphics[width = 0.30 \textwidth]{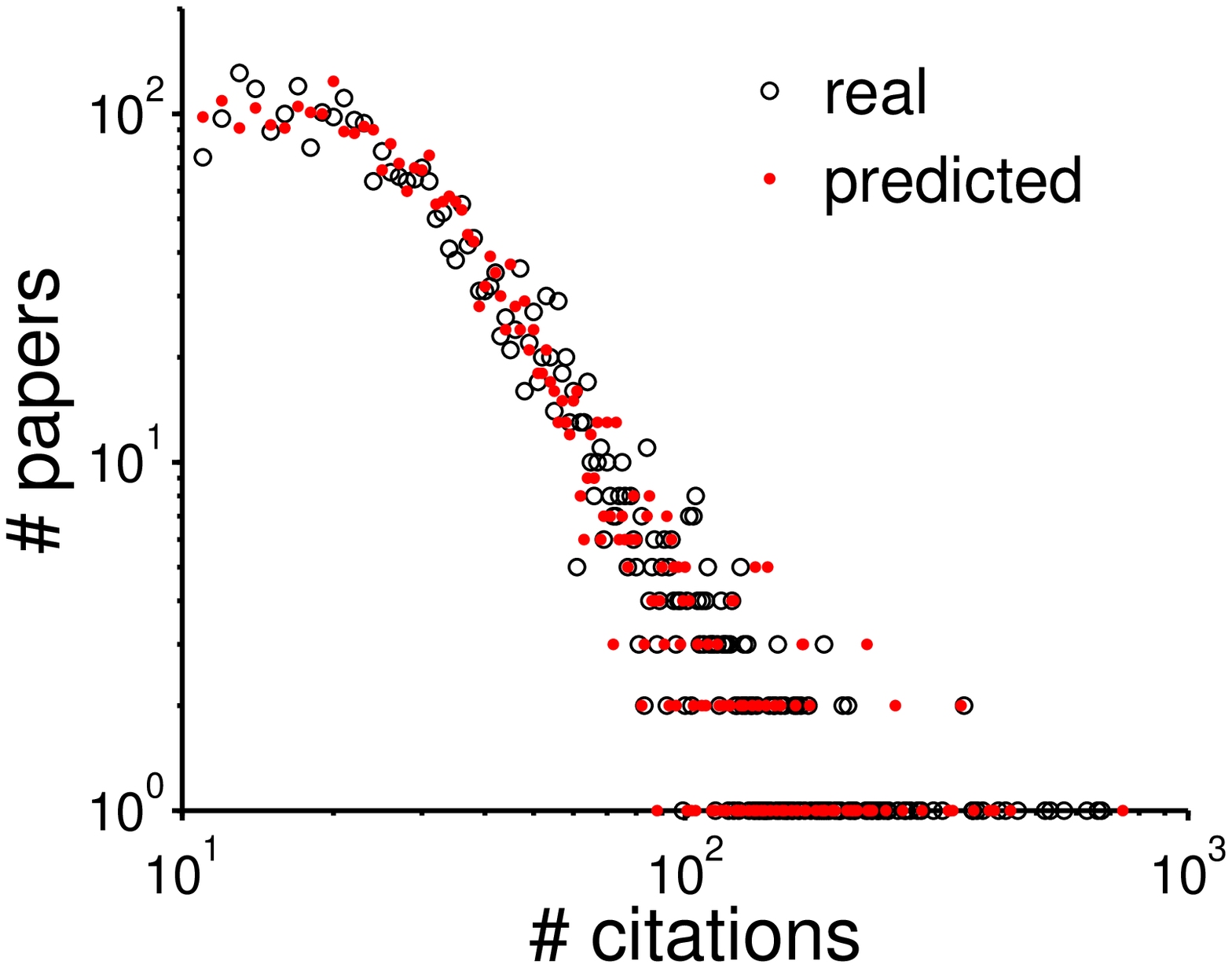}
}
\subfigure[Physical Review Letters (1970s)]{
    \label{fig3h}
    \includegraphics[width = 0.30 \textwidth]{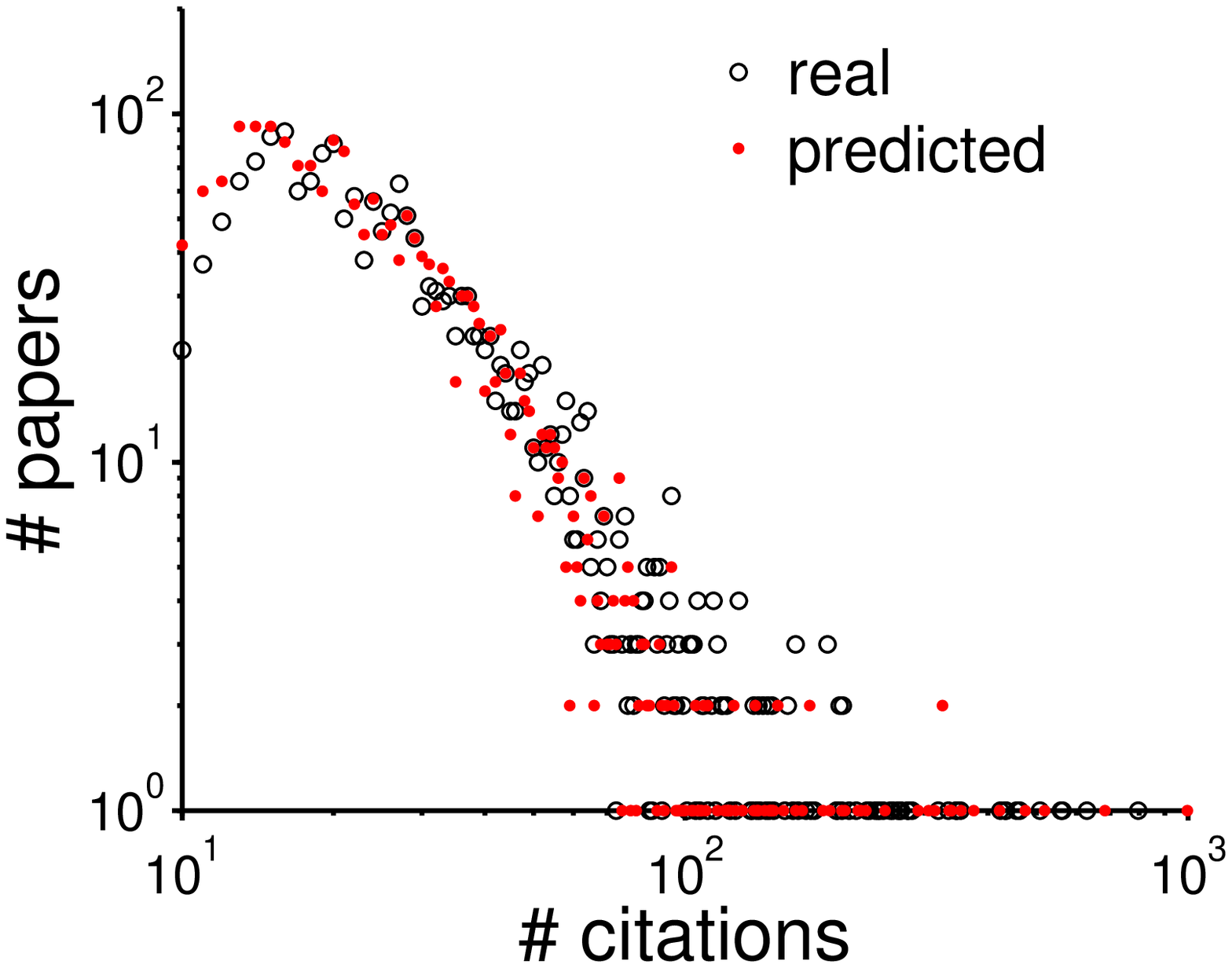}
}
\subfigure[Physical Review B (1980s)]{
    \label{fig3i}
    \includegraphics[width = 0.30 \textwidth]{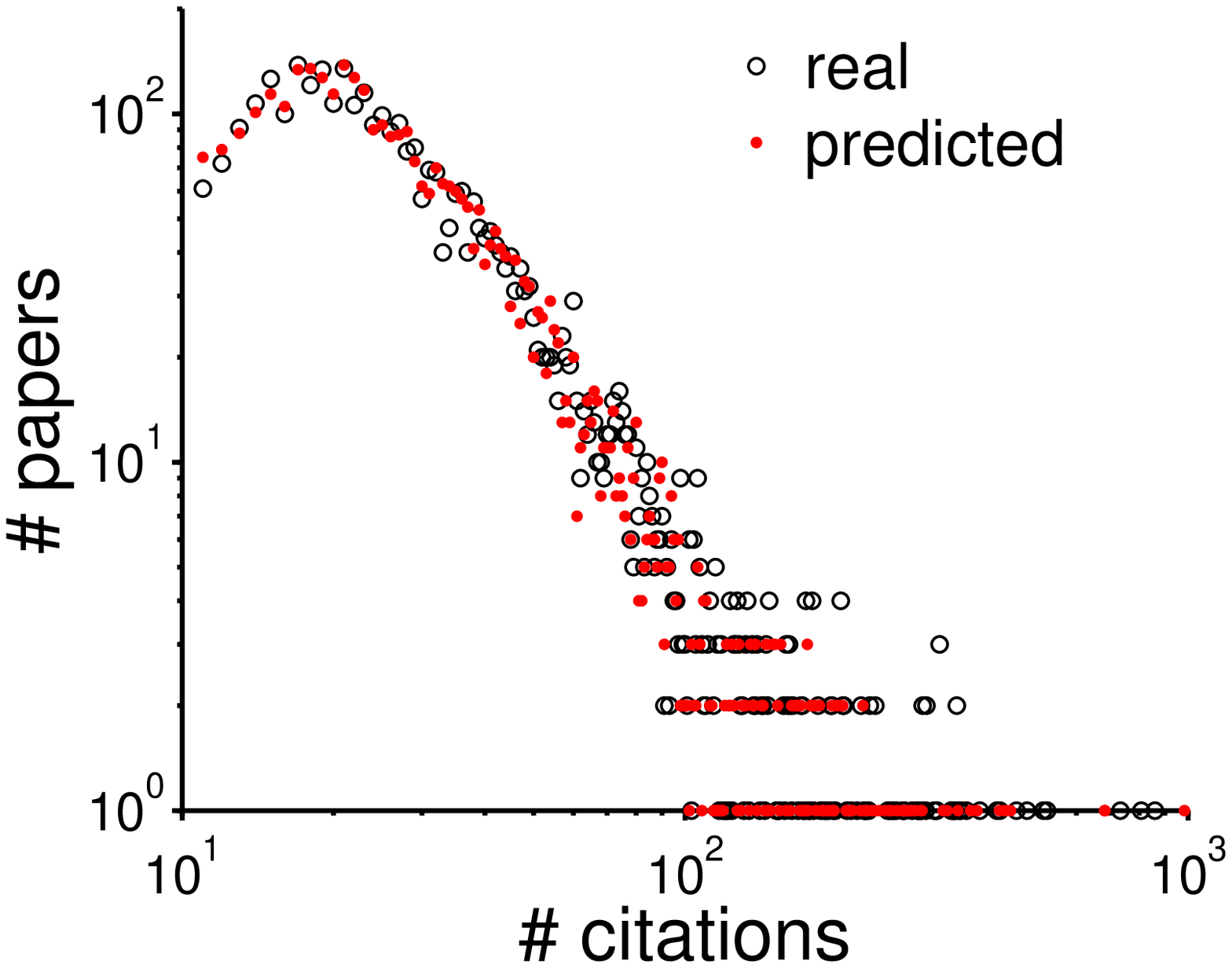}
}
\caption{\label{fig3} The performance comparison in popularity prediction.}
\end{center}
\vskip -0.2in
\end{figure*}

\subsection{Experimental Results}

In this section, we report two sets of experiments. (1) We compare the predictive power of RPP model with other competing methods, finding that RPP consistently outperforms other models in terms of both average deviations and the fraction of papers correctly predicted. (2) We further perform detailed analysis to understand the factors that could affect the performance of RPP model, including the length of training period, the effective number of attention, and the prior parameters.

\noindent \textbf{Popularity prediction.} We evaluate the prediction results on three collections of papers: (a) papers published in Physical Review (PR) from 1960 to 1969; (b) papers published in Physical Review Letters (PRL) from 1970 to 1979; (c) papers published in Physical Review B (PRB) from 1980 to 1989. These samples vary in timeframes and scopes, spanning three decades and covering three types of journals. Among them, in the studied period, Physical Review published articles from all fields of physics. Physical Review Letters published letters (statistically high impact papers) from all fields of physics. Physical Review B published articles from a specific field of physics, i.e., condensed matter physics. Using papers with more than $10$ citations during the first five years after publication, we compare the RPP model with and without prior against the AR and SH models. The number of papers in the three collections is $3242$, $2017$ and $3732$, respectively. The training period is $10$ years and we predict the citation counts for each paper from the 1st to $20$th year after the training period. For collection (c), we predict the citation counts up to the $10$th year after training period due to the cutoff year (2009). We set the parameter $m=30$ for now, corresponding to the typical number of references for a paper, leaving the effect of varying $m$ on the performance of RPP model for later discussions.

We find the RPP model, proposed in this paper, achieves higher accuracy than the AR and SH models (Figure~\ref{fig3}). Yet in absence of prior it only exhibits modest performance in terms of MAPE, indicating that the RPP model without prior performs well on most papers but has rather large errors on a handful of papers. This is caused by its exponential dependence on the fitness parameter that sometimes yields overfitting problem through maximum likelihood parameter estimation. This problem is nicely avoided by incorporating conjugate prior for the fitness parameter, documented by the fact that the RPP model with prior consistently outperforms the other three methods on all collections.

\begin{figure}[t]
\vskip 0.1in
\begin{center}
\centerline{\includegraphics[width=0.85\columnwidth]{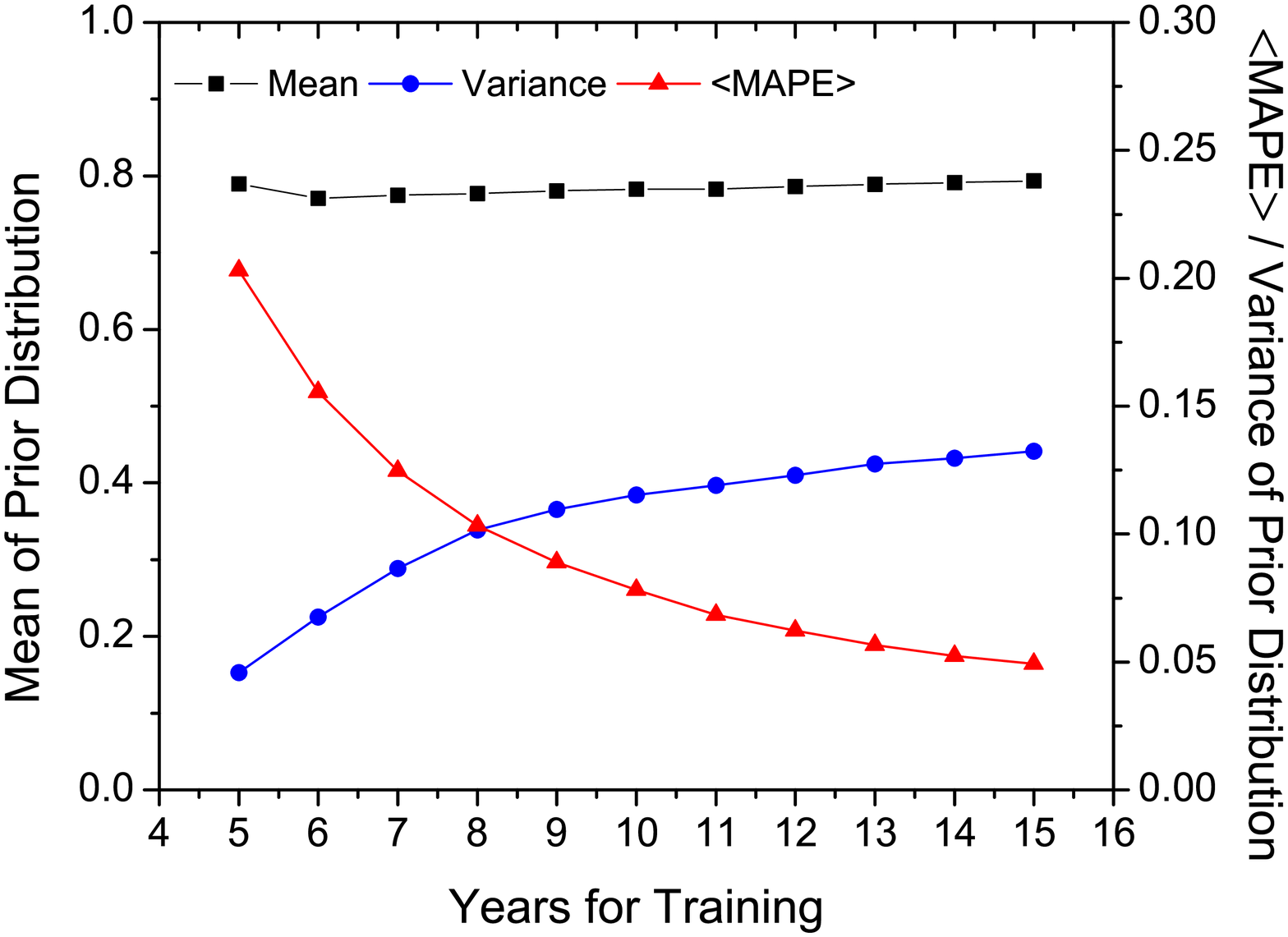}}
\caption{Effect of training period length.}
\label{fig4}
\end{center}
\vskip -0.2in
\end{figure}

The superiority of the RPP model with prior, compared to the AR and SH models, increases with the number of years after the training period. This improvement is rooted in the methodological advantage: the RPP model is a generative probabilistic model that models explicitly the arrival process of attentions, while the two baseline models only capture the correlation between early popularity and future popularity, no matter linearly or logarithmically. In addition, the reinforced Poisson process could model the ``rich-get-richer'' phenomenon in popularity dynamics and thus could characterize the logarithmic correlation between early popularity and future popularity. Therefore, when compared with the AR method, the superiority is more obvious than being compared with the SH method. This is because the AR method works linearly while the SH method works in a logarithmic manner.

The RPP models with and without prior are trained only on the popularity dynamics during training period while the training of the AR and SH models depends on the knowledge of future popularity dynamics. When training these two models, we employ the leave-one-out technique which uses all papers except the target paper for prediction. Yet, in most cases, it is unrealistic to know future popularity dynamics when training the model, limiting their  applications in real scenarios.

Finally, being a generative model, the RPP model is able to reproduce the citation distribution. Indeed, as shown in Figure~\ref{fig3} (g-i), the distribution of citations predicted by the RPP model with prior matches very well with that of real citations on all studied collections, indicating that the RPP model can also be used to model the global properties of citation system.

\noindent \textbf{Analysis of relevant factors.} The superior predictive power in the RPP model with prior raises an interesting question: what are the possible factors that affect its predictive power? In this section, we study a number of factors which could affect the performance of the RPP model with prior. Hereafter, we use $\langle$MAPE$\rangle$ to denote the average MAPEs for predictions from the 1st to 10th year after training period. The training period is 10 years except when we discuss the effect of varying training period length. The parameter $m$ is set to be $30$ except when we discuss the effect of changing $m$.

First, we study the prediction accuracy of the RPP model with prior by varying training period. Experiments are conducted on the collection of papers published in Physical Review from 1960 to 1969. As shown in Figure~\ref{fig4}, $\langle$MAPE$\rangle$ decreases as the training period increases. Hence increasing the training period improves the prediction accuracy. However, the rate at which $\langle$MAPE$\rangle$ diminishes slows down quickly, indicating the marginal gain of increasing training period. We also find that the mean of prior distribution stays almost constant as the length of training period increases from 5 years to 15 years, indicating the expected fitness parameter learned by the RPP model is robust against varying training period. At the same time, a longer training period could reduce the role of prior in prediction, partly explaining the role of prior in overcoming the overfitting problem, as demonstrated by the increasing variance in the prior distributions with the length of training period.

\begin{table}[t]
\caption{Effect of the number of conceived attention.}
\label{tab2}
\vskip 0.1in
\begin{center}
\begin{tabular}{lccc}
\hline
$m$ & Mean ($\alpha/\beta$) & Variance ($\alpha/\beta^2$) & $\langle$MAPE$ \rangle$ \\
\hline
10  &   1.467 & 0.193 & 0.0762 \\
20  &   1.005 & 0.150 & 0.0776 \\
30  &   0.783 & 0.115 & 0.0781 \\
40  &   0.647 & 0.091 & 0.0784 \\
50  &   0.554 & 0.074 & 0.0785 \\
\hline
\end{tabular}
\end{center}
\vskip -0.15in
\end{table}

Second, we investigate the effect of parameter $m$, i.e., the effective number of attention by conducting experiments on the paper collection (a). Intuitively, $m$ balances the strength in the reinforcement mechanism. Indeed, as shown in Table~\ref{tab2}, the mean and variance of the prior distribution decay with $m$, demonstrating these parameters are mainly determined by papers with fewer citations. We also find that decreasing $m$ reduces $\langle$MAPE$\rangle$, indicating that the disparity in citations is captured appropriately by the reinforcement mechanism in our model, as a larger $m$ implies a weaker role of the reinforcement mechanism. Token together, Table~\ref{tab2} confirms that the reinforcement mechanism is crucial to modeling popularity dynamics in citation system.

\begin{table}[t]
\caption{Prediction accuracy over four decades.}
\label{tab3}
\vskip 0.1in
\begin{center}
\begin{tabular}{lcccc}
\hline
Period & $\alpha$ & $\beta$ & $\alpha/\beta$ & $\langle$MAPE$\rangle$ \\
\hline
1950s  &   4.237 &  4.061 & 1.043 & 0.075 \\
1960s  &   4.759 &  4.440 & 1.072 & 0.084 \\
1970s  &   6.130 &  4.924 & 1.245 & 0.111 \\
1980s  &  10.706 &  5.379 & 1.990 & 0.120 \\
\hline
\end{tabular}
\end{center}
\vskip -0.15in
\end{table}

Finally, we use papers published in Reviews of Modern Physics (RMP) to illustrate the change of prior parameter $\alpha$ and $\beta$ over four decades and their influence on the prediction accuracy of the RPP model with prior. As shown in Table~\ref{tab3}, the mean of prior distribution (i.e., $\alpha/\beta$) increases with the increasing magnitude of both $\alpha$ and $\beta$ over the four decades. This indicates that the expected citations for papers in this prestigious journal steadily increases in the second half of the 20th century. Meanwhile, the $\langle$MAPE$\rangle$ of the RPP model also increases. Hence it becomes more difficult to predict the citations of these papers, as a result of the increasingly skewed distribution of citations.

\section{Related Work}

Modeling and Predicting popularity dynamics is a fundamental problem in different areas, including the diffusion of innovation, social contagion, information propagation, and other social dynamics. Existing studies on popularity dynamics include influence spread~\cite{Kempe2003}, trust propagation~\cite{Guha2004}, information access pattern~\cite{Dezso2006}, group formation~\cite{Backstrom2006}, culture market~\cite{Salganik2006}, popularity prediction of online content~\cite{Crane2008,Szabo2010}, Web users' behavior~\cite{Lerman2012}, and citation prediction~\cite{Yan2011,Yu2012}. These studies focus mostly on analyzing the factors and mechanisms affecting popularity dynamics, such as structural and temporal patterns. Few of these approaches model how individual item accrues its popularity, which is the focus in this paper.

Empirical studies show that attentions are allocated in a rather asymmetric way, i.e., most items receive little attentions whereas a few acquire a disproportionately large fraction of the total attentions~\cite{Szabo2010}. The inhomogeneous attention distribution, as a whole, has been well understood as a consequence of collective human behavior~\cite{Barabasi1999,Barabasi2005,Ratkiewicz2010}. However, it is largely unclear to model and predict the popularity of individual items.

The key of popularity prediction is analyzing and understanding the underlying dynamics, which characterizes the evolution of popularity over time. It is widely believed that the popularity dynamics is governed by
users' collective actions~\cite{Crane2008}. Most existing approaches on popularity prediction treat popularity dynamics as a time series and predict future popularity by exploiting certain temporal patterns and correlations of these time series~\cite{Szabo2010,Yang2010}. Furthermore, intrinsic attractiveness of item and the underlying social network structure are explored and incorporated into the time series analysis to improve the predictive power of these methodology~\cite{Lerman2010,Bao2013b}. While previous models provide some insights about temporal and structural patterns in popularity dynamics, they fail to model directly the arrival process of attentions.

Several works attempt to model popularity dynamics using traditional models for epidemic spread and diffusion of innovation~\cite{Crane2008,Ugander2012}. These models are essentially descriptive models and lack predictive power. Recently, reaction-diffusion models and branching stochastic processes were adopted to characterize popularity dynamics. Among them, the so-called self-excited Hawkes conditional Poisson process has been used successfully to model the power-law relaxation of popularity dynamics and represents a promising predictive power~\cite{Matsubara2012}. However, the dependence on exogenous factors and the hard-coded power-law relaxation function limit its applicability to specific contexts. Alternatively, survival theory is used to model information propagation and to predict the size of information cascades~\cite{Gomez-Rodriguez2013}. However, this model only considers the arrival time of attention and the time interval between successive arrivals of attention, and thus fails to characterize the ``rich-get-richer'' phenomenon observed in popularity dynamics.

\section{Conclusions}

Taken together, we presented a general framework to model and predict popularity dynamics based on a reinforced Poisson process. This model incorporates three key ingredients of popularity dynamics: the fitness parameter characterizing intrinsic attractiveness, the temporal relaxation function explaining the aging effect in attracting new attentions, and the reinforcement mechanism corresponding to the ``rich-get-richer'' effect in popularity dynamics. Being a generative probabilistic framework, it models explicitly the stochastic process of gaining popularity for each item, in direct contrast to existing deterministic approaches. We developed optimization methods to train the proposed RPP model with and without priors. The RPP model with prior allows us to apply the Bayesian treatment, resulting in more robust and accurate predictions for popularity dynamics. We empirically validate our model on an excellent longitudinal dataset on citations, spanning over one hundred years, demonstrating its clear advantages over competing methods.

The model's flexibility in its relaxation function makes it a general framework to model popularity dynamics that can be adapted in different domains. While previous works suggest the log-normal function works well for citation dynamics~\cite{Wang2013} or the power-law function is better suited for online videos~\cite{Crane2008}, a more systematic framework to identify such functions for a specific domain would be a promising future direction.
Another interesting direction is to explore ways to enrich the proposed model by incorporating relevant factors within each specific domain, and the improvement enabled by these factors in both prediction accuracy and shortened training period could shed new light on the nature of popularity itself. Hence, being a general framework, the proposed model offers a springboard to anchor and benchmark future models, and is expected to play an increasingly important role as new and increasingly detailed data flourish and our understanding of quantitative laws behind popularity dynamics deepens.

\bibliographystyle{abbrv}
\bibliography{rpp}

\end{document}